\documentclass[%
 aps,reprint,
showpacs,
 amsmath,amssymb,
 prl,
]{revtex4-1}

\pdfoutput=1
\usepackage{graphicx}
\usepackage{dcolumn}
\usepackage{bm}
\usepackage{tikz}
\usetikzlibrary{arrows}
\usepackage{wrapfig}
\usepackage{amssymb}
\usepackage{epstopdf}
\usepackage{float}
\usepackage{xcolor}
\usepackage[top=0.75in, bottom=0.75in, left=0.5in, right=0.5in]{geometry}
\usepackage{enumitem}
\setlist[enumerate]{itemsep=0mm}
\usepackage{fancyhdr}
\usepackage{mathtools}

\usepackage{color}

\usepackage[mode=buildnew]{standalone}

\graphicspath{{fig/}}

\begin{document}

%

\title{Binary Operations on Neuromorphic Hardware with Application to \\Linear Algebraic Operations and Stochastic Equations}
\author{Oleksandr Iaroshenko}
\author{Andrew T. Sornborger}\email[Corresponding author's email: ]{sornborg@lanl.gov}
\affiliation{$^1$Information Sciences, CCS-3, Los Alamos National Laboratory, Los Alamos, NM 87545, USA}

\date{\today}

\begin{abstract}
\noindent 
Non-von Neumann computational hardware, based on neuron-inspired, non-linear elements connected via linear, weighted synapses -- so-called neuromorphic systems -- is a viable computational substrate. Since neuromorphic systems have been shown to use less power than CPUs for many applications, they are of potential use in autonomous systems such as robots, drones, and satellites, for which power resources are at a premium. The power used by neuromorphic systems is approximately proportional to the number of spiking events produced by neurons on-chip. However, typical information encoding on these chips is in the form of firing rates that unarily encode information. That is, the number of spikes generated by a neuron is meant to be proportional to an encoded value used in a computation or algorithm. Unary encoding is less efficient (produces more spikes) than binary encoding. For this reason, here we present neuromorphic computational mechanisms for implementing binary two's complement operations. We use the mechanisms to construct a neuromorphic, binary matrix multiplication algorithm that may be used as a primitive for linear differential equation integration, deep networks, and other standard calculations. We also construct a random walk circuit and apply it in Brownian motion simulations. We study how both algorithms scale in circuit size and iteration time.
\end{abstract}


\maketitle


\section{Introduction}
\noindent
In the context of neural systems, where, due to the expense of synaptic transmission, a spike costs more than the absence of a spike, sparse communication is more efficient. This fact has been studied both in biological neural systems \cite{harris2012synaptic,levy2002energy} as well as in neuromorphic systems \cite{daviesSpikesAI}. 

However, biological neural computations as well as algorithms implemented on neuromorphic chips typically use some form of unary coding, i.e. information is encoded in firing rates. Though energy efficient at low spiking rates, unary coding can be wasteful in terms of the number of spikes used for a given numerical resolution. Each wasted spike contributes excess power used by the neural circuit. 

A previous attempt to use more efficient spike coding in a neuromorphic machine learning algorithm used a mixed binary-unary encoding to make multiplication more efficient \cite{DiehlEtAl2016}. Here, we present three implementations of circuits using binary operations: 1) a fully binary implementation of multiplication, where a two's complement number encoded in spikes multiplies a two's complement number encoded in synapses, 2) an extension of this circuit to vector-matrix multiplication, and 3) a simpler, and highly parallelizable, circuit for performing a random walk. 

Our neural circuits use a pulse-gating framework to control and coordinate the flow of spikes between a number of cores. They are made up of five neuron types and standard synaptic connections are used for the implementation.

By encoding numbers in binary, we logarithmically reduce the number of spikes needed to represent a binary-encoded number or vector. This reduction comes at the cost of increasing the neurosynaptic footprint needed to represent the neural circuit. 

Matrix multiplication has a broad range of uses across the spectrum of computational algorithms. The neural circuits that we develop below may be used for various applications, including implementing kernels for machine learning to more prosaic computational algorithms. Here, we consider 1) simulations of linear dynamics, and 2) simulations of Brownian motion. The first application is straightforward with the availability of the matrix-vector multiplication circuit that we demonstrate. We implement the second application in an easily parallelizable neural circuit. 

Both of the applications that we consider have been studied in neuromorphic circuits previously \cite{severa2018spiking,parekh2017constant,reeder2019future}. The circuits we present here use two's complement binary operations and scale poly-logarithmically in the number of neurons as a function of numerical resolution, whereas the modular spatial code using a residue numerical system presented in \cite{severa2018spiking} offers a compromise between unary and binary codings with intermediate scaling. The time scaling for our circuits is also poly-logarithmic in the numerical resolution, whereas with unary encoding, scaling would be linear in the numerical resolution (i.e. exponentially more expensive than binary). The neural circuits that we present here should prove useful in providing low-spiking rate, low-power algorithms for linear algebraic applications in neuromorphic circuits.

\section{Methods}
\subsection{Neural dynamics}
We implemented our two's complement algorithms on Intel’s neuromorphic research processor codenamed Loihi \cite{DBLP:journals/micro/DaviesSLCCCDJIJ18}.
A single Loihi chip consists of $128$ logical cores with each core having $1024$ compartments and an upper bound of $2^{14} = 16,384$ synapses per core.
A simple neuron occupies one compartment whereas more sophisticated neurons can be built from several compartments.
The neural dynamics is computed at discrete timesteps called ticks. 

For our binary multiplication circuit, we make use of three types of simple neurons:
\begin{itemize}
    \item $\gamma_0$ -- a TRANSFER neuron, which emits a spike one tick after it receives an incoming spike from any of its inputs;
    \item $\gamma_1$ -- a GATING neuron, which emits a spike one tick after receiving at least two incoming spikes at the same tick;
    \item $\gamma_2$ -- a COUNTER neuron, which emits a number of spikes equal to the number of incoming spikes at time $t_0$, at a rate of one spike per tick starting from time $t_0+1$. 
\end{itemize}

Loihi's neuronal compartments convert incoming spikes to a current which is multiplied by a synaptic weight and accumulated as an internal voltage:
\begin{equation}\label{eq:neuro_dynamics}
    \begin{cases}
        u_i(t) = u_i(t - 1)\cdot(2^{12} - \delta_i^{(u)})\cdot 2^{-12} + 2^6\cdot\sum_j w_{ij} s_j(t), \\
        v_i(t) = v_i(t - 1)\cdot(2^{12} - \delta_i^{(v)})\cdot 2^{-12} + u_i(t) + u_j^{\textup{bias}}.
    \end{cases}
\end{equation}
Here $i$ and $j$ are the indices of the receiving and emitting neurons (resp.), $u$ is the current in the receiving neuron, $v$ is the voltage in the receiving neuron. $\delta_i^{(u,v)} \in [0, 2^{12}]$ are current and voltage decay constants respectively, $w_{ij}$ is the synaptic weight, $s_j(t) \in \{0, 1\}$ are upstream emitted spikes, and $u_j^{\textup{bias}}$ is a constant bias current.
The variable $t \in \mathbb{N}$ represents time (in units of ticks).

We use the same synaptic weight throughout, $w_{ij} \equiv w = 100$, and a single spike current $u_s = 2^6$ (by default in the Loihi API) as shown in equation \eqref{eq:neuro_dynamics}.
When the voltage exceeds a threshold, $v_{\textup{th}}$, a spike is emitted and the voltage is reset to zero.
Both the incoming current and the voltage can have their own exponential decay rules defined by the decay constants $\delta_i^{(u,v)}$ and the system of equations \eqref{eq:neuro_dynamics}.
In all neurons $\gamma_j$, $j=0,1,2$, the voltage decay is immediate ($\delta_i^{(v)} = 2^{12}$) and resets to zero at every tick.
The current decay is immediate for $\gamma_0$ and $\gamma_1$ but is linear for the spike counter, $\gamma_2$, which we explain below.

With this modeling framework, the TRANSFER and GATING neurons are straightforward to model.
The TRANSFER neuron has a voltage threshold just below the product of spike current and synaptic weight, $v_{\text{th}} = u_s w - 1$. So, as soon as it receives an input spike, the voltage jumps over threshold and emits a spike at the next tick.
The GATING neuron has a voltage threshold above the spike current-weight product, $v_{\text{th}} = u_s w + 1$, but below twice this value. Thus, it does not respond to a single spike but emits a spike if it receives two (or more) spikes.
If a GATING neuron has only 2 dendrites, this behavior is equivalent to a logical AND operation and hence we use it to gate the transfer of spiking information dependent on whether a {\it gating} spike and an {\it information} spike arrive simultaneously at the dendrites of the GATING neuron.

The COUNTER neuron, $\gamma_2$, uses current decay to drive the continuous emission of spikes during a period of time equal to the current-weight product of a set of input spikes to a set of dendrites.
The current decay per tick equals a single spike's current. Therefore, the incoming current persists for a time period equal to the current-weight product received at a given tick then drops linearly to zero.
This neuron works correctly (i.e. as a neuron to count out a current-spike product as a set of spikes) if it does not receive any spikes while its incoming current is decaying to zero.
To model COUNTER neuron behavior we set the current decay to zero ($\delta_i^{(u)} = 0$) but connect the neuron to itself with an inhibitory synapse.
Therefore when it emits a spike, it sends itself a negative current, equivalent to subtracting a single spike's current.
In this way, the current decreases one spike current per tick while emitting spikes until its current value drops to zero.

\subsection{Two's Complement Multiplication}
In order to multiply two $M$-bit integers in two's complement notation, we first extend them to $N = 2M$-bit numbers.
By {\it extend} here, we mean duplicating the highest bit to fill out the additional bits. For example, $1011 \rightarrow 11111011$ or $0011 \rightarrow 00000011$.
Input to our binary multiplication circuit is a vector of spikes (the multiplier), where existence of a spike indicates a $1$ and non-existence of a spike indicates a $0$. The lowest bit is input on an axon representing $2^0$ and the highest bit is input on an axon representing $2^{N-1}$, and similarly with axons in between. 

The vector of spikes propagates through a synaptic connectivity (i.e. a matrix) that encodes a binary number (the multiplicand). This encoding is done by shifting the binary representation (synapse of $1$ represents binary digit of $1$ and synapse of $0$ represents binary digit of $0$) of the multiplicand across the set of axonal synapses (see Fig.~\ref{fig:multiply} for an encoding of the two's complement number $-1$). Shifts are performed with truncation at the highest bit.

Our multiply subcircuit (core) uses COUNTER neurons. Therefore, when the multiplier and multiplicand are multiplied, the multiply operation outputs a sequence of at most $N$ sets of spikes representing a mixed binary-unary representation.

The rest of the circuit is dedicated to sorting this mixed representation such that pairs of numbers output by the multiply circuit may be carried and added, then recombined, with subsequent carries and adds performed iteratively until all carries are performed and only a binary representation remains.

\subsection{Constituent Cores Making up the Full Circuit}
Our full circuit is composed of six different types of core.
The purpose of these cores is 1) to turn the circuit on or off; 2) to control circuit timing and to control how information is distributed across the circuit; 3) to multiply input spikes representing the multiplier by synaptic weights representing the multiplicand; 4) to spatially distribute the resulting spikes; 5) to perform carry/add operations on pairs of spike registers; and 6) to aggregate intermediate results and iterate the computation to completion.

\begin{figure}
\includegraphics[width=0.55\columnwidth]{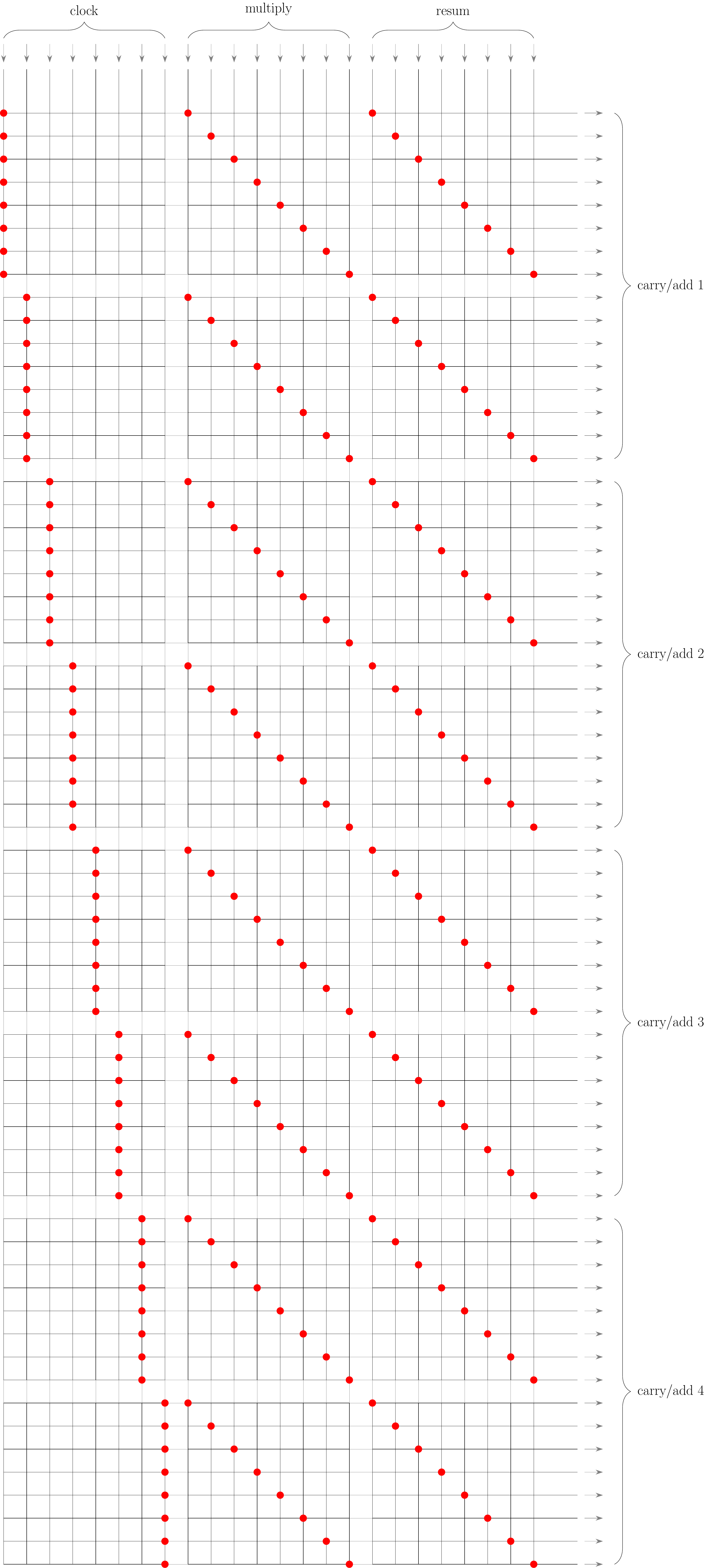}
\caption{The Multiplexer Core. Gating spikes from the clock (left register) enable through-propagation of spikes arriving either from the multiply core or from the resum core.}
\label{fig:multiplexer}
\end{figure}

\begin{figure}
\includegraphics[width=0.95\columnwidth]{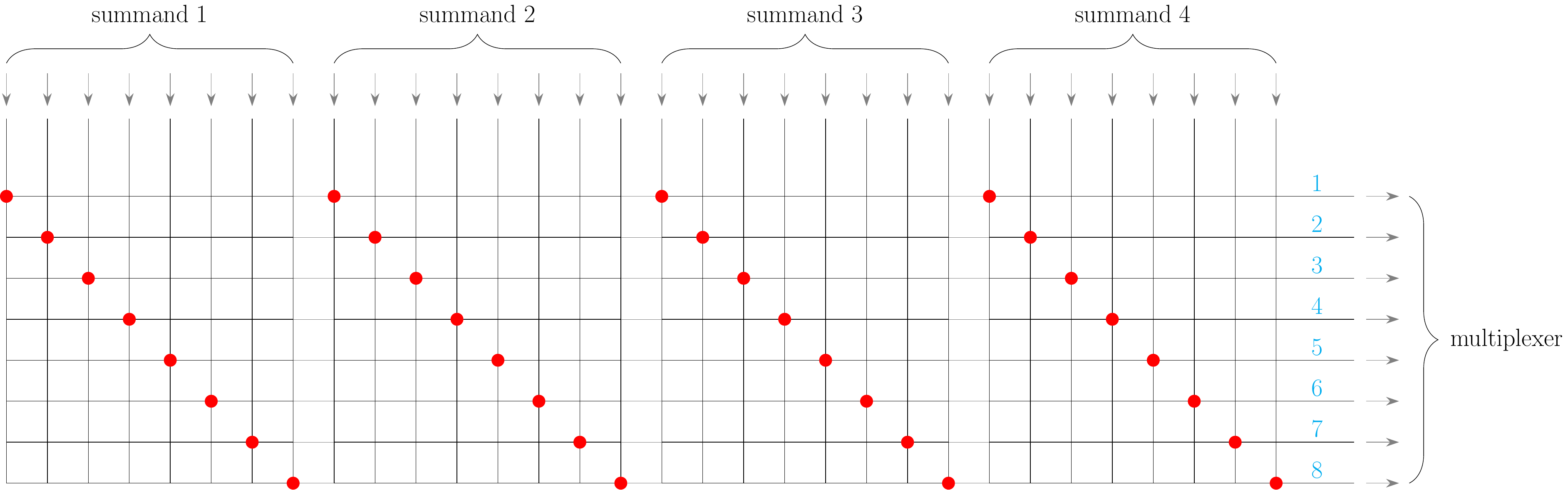}
\caption{The Resum Core. Consisting of COUNTER neurons, this core adds input spikes and outputs a new mixed unary-binary representation that is sent to the multiplexer for a new carry-add iteration.}
\label{fig:resum}
\end{figure}

These tasks are performed by on/off, clock, multiply, multiplexer, carry/add, and resum cores (resp.), whose functions are outlined below (see Figs.~\ref{fig:clock}-\ref{fig:resum}).
We present example cores used for multiplying two $4$-bit numbers.
The generalization to an arbitrary number of bits is straightforward.

%
%

\subsubsection{On/Off Core}
The on/off core (Fig.~\ref{fig:clock}) has two neurons responsible for turning the circuit on or off.
This is implemented using a connection to the clock core.
The ON neuron sends a spike to the first neuron in the clock core.
It starts the clock and the incoming spike starts circulating inside the clock core.
A spike from the OFF neuron inhibits all activity in the clock core and stops the clock.

\subsubsection{Clock Core}
\begin{figure}
\includegraphics[width=0.95\columnwidth]{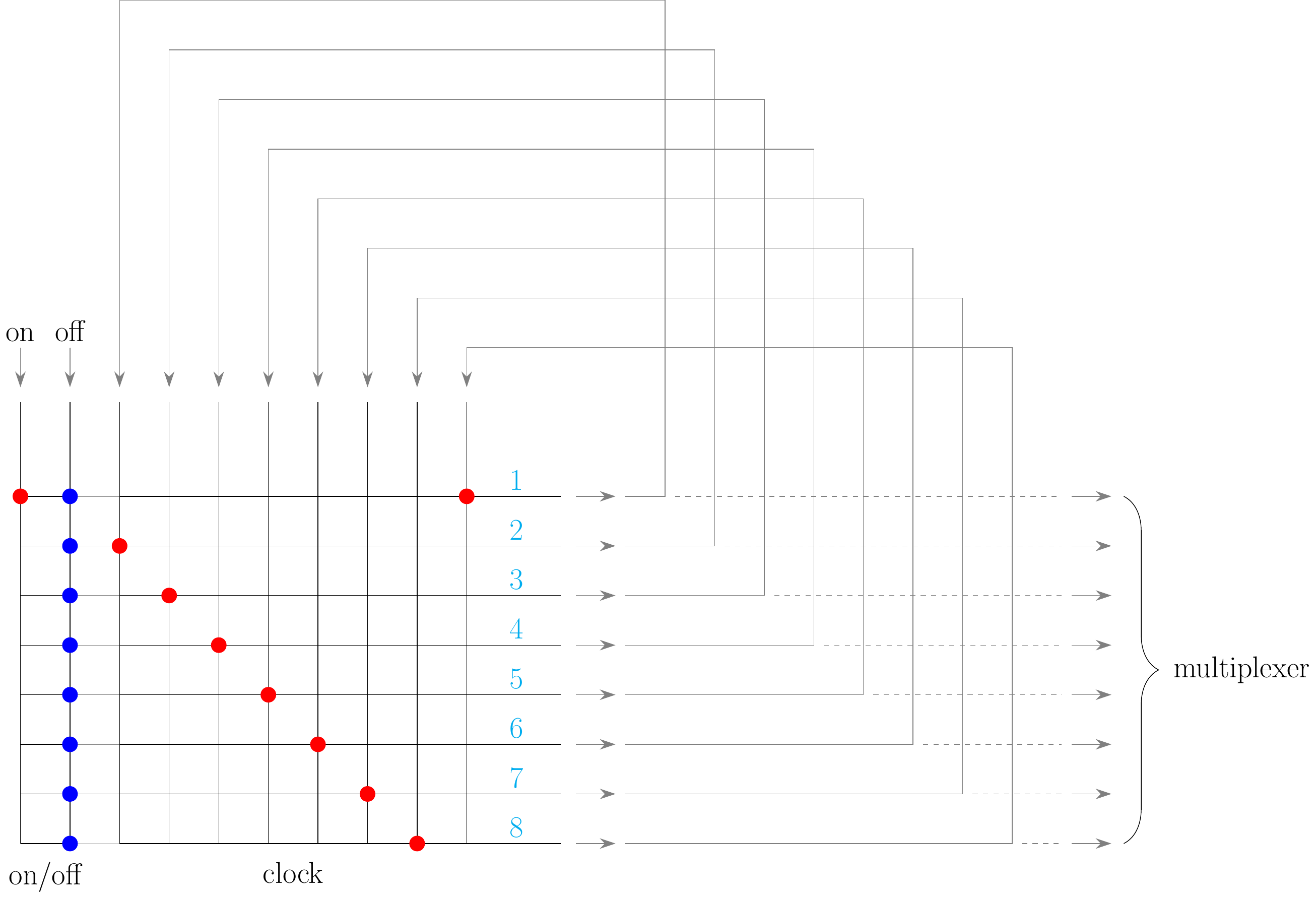}
\caption{The On/Off and Clock Cores. Here, and in figures below, inputs arrive from the top and outputs are emitted to the right. {\it on} and {\it off} inputs to the clock core (left two columns) are indicated at top left. A cyclic circuit (clock core) implementing an $N$-tick clocking cycle is the main set of axons and neurons on the right. Recurrent connections implementing the cycle are shown to the right. Output is additionally sent downstream to the multiplexer circuit. (See text for description.)}
\label{fig:clock}
\end{figure}
The clock core (Fig.~\ref{fig:clock}) has $N$ $\gamma_0$ neurons which indicate the state of an $N$-tick clocking cycle.
After the first neuron receives a spike, it emits the spike to the multiplexer (below) and also sends the spike to the second neuron.
The second neuron emits the spike to the multiplexer and sends it to the third neuron and so on.
The last neuron sends the spike to the first neuron after emitting it to the multiplexer.
The emitted spikes communicate the clock state to the multiplexer core, and serve as gating spikes to regulate the multiplexer core's function.

\subsubsection{Multiply Core}
The multiply core is designed to calculate the partial products of the multiplicand, that are encoded in synaptic weights, by each bit of the multiplier, which are encoded in spikes input to the multiply core.
Basically, this operation mimics the standard two's compliment multiplication algorithm as it would be performed manually, where the partial products of the multiplier and multiplicand are calculated and then added together.
In order to add multiple binary partial products we use a combination of carry/add, multiplexer, and resum cores.
We use $\gamma_2$ neurons for the multiply core, in which each neuron $i$ will emit $k_i$ spikes during $k_i$ ticks, where $k_i$ is the summed current-weight input to neuron $i$.

\begin{figure}
\includegraphics[width=0.95\columnwidth]{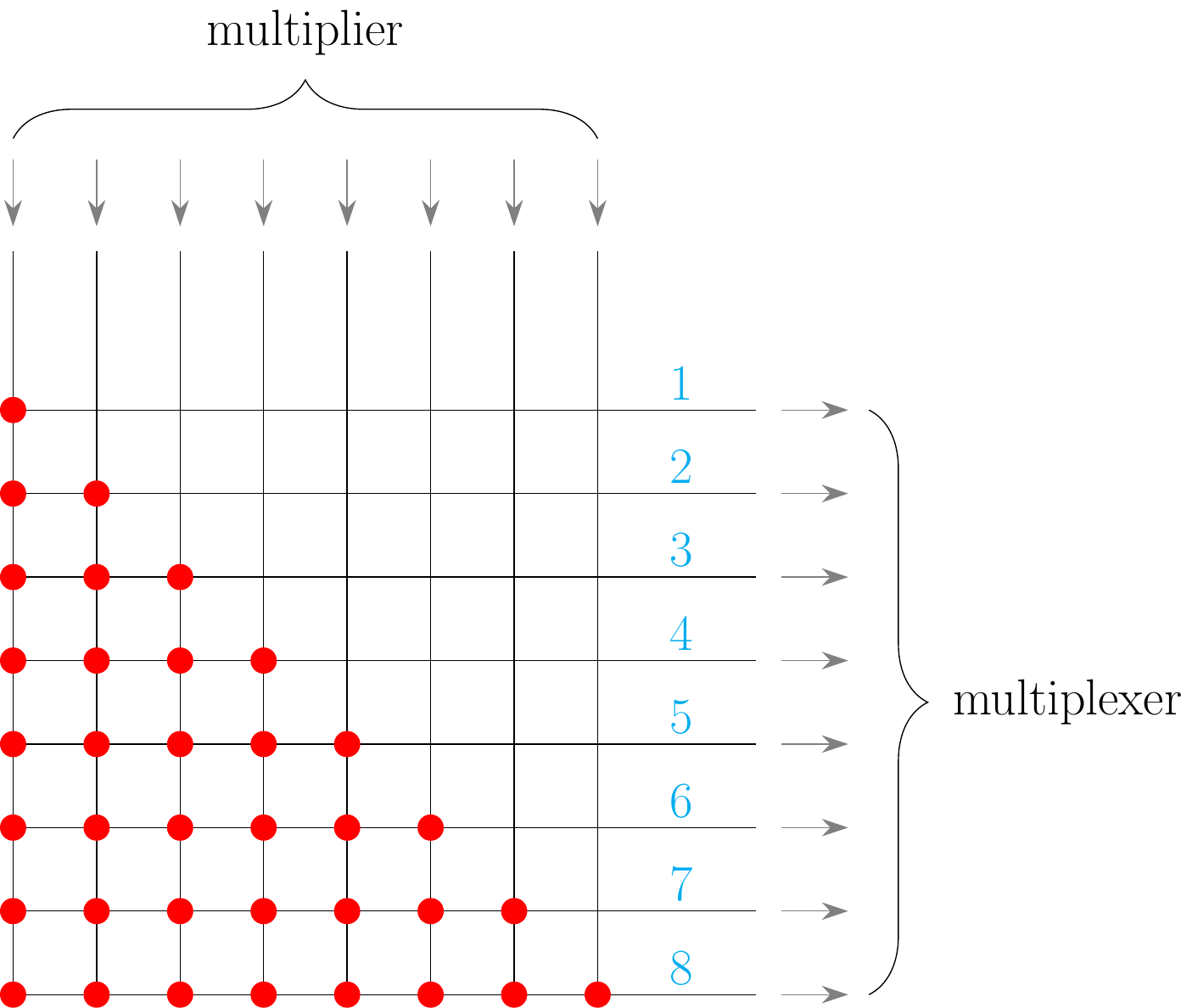}
\caption{The Multiply Core. The two's complement representation of $-1 = 11111111$ is shown encoded in the synapse in this example. This representation is shifted and truncated across the set of inputs. All neurons are COUNTER neurons in this core. See text for further details.}
\label{fig:multiply}
\end{figure}

Figure \ref{fig:multiply} shows the positive synapses that represent the number $-1$ in $8$-bit two's complement notation (i.e. $11111111$).
For each subsequent spiking input from the multiplier to the multiply core the binary representation of the multiplicand is shifted to accommodate for the different positions of the bits in the multiplier.
For example, if the multiplier is binary $11111111$, then the spikes come from all input channels.
E.g. the first neuron receives a single spike at time $t_0$ and emits a spike at time $t_0+1$.
In turn the second neuron receives two spikes at time $t_0$ and emits spikes during two ticks from $t_0+1$ to $t_0+2$.
In general if the multiplier and the multiplicand are $a$ and $b$ respectively, then the spike from channel $i$ (which means that $a$ has $1$ at the $i$th position in two's complement notation) leads to the output that encodes $2^i \cdot b$ which is $b$, shifted by $i$ positions.
{\color{black} For this 4-bit implementation, the shifted multiplicand is truncated at 8-bits, since carries beyond this bit are unnecessary.}

\subsubsection{Multiplexer Core}
The role of the multiplexer is to provide the correct input for the carry-add cores from multiply and resum cores based on timed gating spikes from the clock.
It has $3 N^2$ neurons that consist of $3 N$ groups of $N$ neurons.
Each group fires only at a corresponding clock state.
This is done in order to synchronize the input from different carry-adds to the resum core.
Since the input is coming from the multiply and the resum cores and they use spike counter neurons, which emit a sequence of spikes during a certain period of time after receiving the input, the clock and synchronization are needed in order for all outputs to combine together at a single tick. 

The input for the multiplexer comes either from the multiply core or the resum core but not both simultaneously since the multiply core shuts down after the first $N$ ticks after receiving input and the resum core does not process any spikes before that.
Both the multiply and the resum cores can keep firing for  at most $N$ ticks after receiving a single input as this is the maximum number of spikes that a single neuron can receive.
Each of $N$ groups in the multiplexer receives spiking input from the multiply and resum cores but receives different gating spikes from the clock depending on the state in its cycle.
{\color{black}The gating spikes will only allow through-propagation of spikes that came from the input core in the $j$th group, where $j\in1,\dots,N$ is the clock state.}

\subsubsection{Carry/Add Core}
The carry/add core receives two binary numbers simultaneously from the multiplexer as an input (one of the inputs is delayed to achieve simultaneity, see Fig.~\ref{fig:circuit}) and processes them using $\gamma_2$ spike counter neurons.
Since the multiplexer outputs $N$ numbers, we need $N/2$ carry-add cores to add them in pairs.
As output the carry/add core emits (at most) two sets of spikes during two ticks after performing a single-bit carry-add operation.
These spikes are transferred to the resum core, which combines them all together for further processing.
\begin{figure}
\includegraphics[width=0.95\columnwidth]{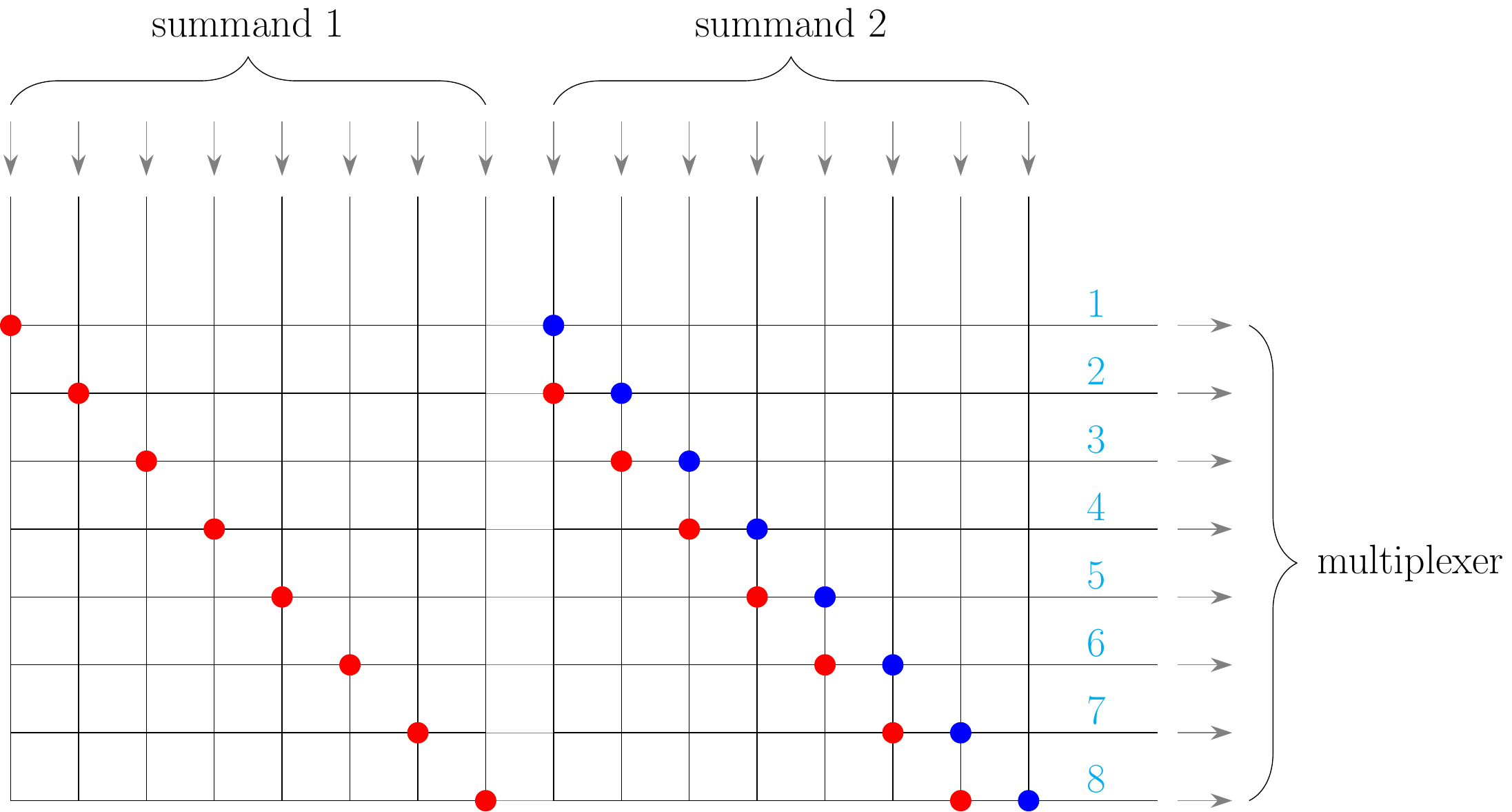}
\caption{The Carry/Add Core. This core combines input from two summands and, for two coincident spikes, performs a carry and adds the spikes. Note that there is an assumption that if only one of the summands has a spike on a particular input, it is summand 1. This assumption may be made since spike sequences from the multiply and resum cores consist of a sequence of all ones, followed by zeros. I.e. there are no zeros interspersed in the spiking sequence.}
\label{fig:carry_add}
\end{figure}
This core operates under the assumption that two binary numbers $a$ and $b$ are coming as a sequential output of a spike counter neuron. This means that if, for a certain position $i$, $a_i$ is zero, then $b_i$ is zero too.
If the $\gamma_2$ neuron did not emit a spike at time $t$, it will not emit it at any further time unless it receives a new input.

The summation scheme is shown on Figure \ref{fig:carry_add}.
If, for a certain position, $b_i = 1$ (then $a_i = 1$, too), then the corresponding neuron receives both positive and negative current values which cancel each other.
If, in addition, the previous bit had two ones ($b_{i-1} = 1$), then that bit carries the value of one to $i$th position and adds a positive current value there.
Thus, only if $a_{i-1} = 1$, $b_{i-1} = 1$, and $a_i = 1$, but $b_i = 0$, will neuron $i$ fire two spikes, otherwise it will fire zero or one spike.

\subsubsection{Resum Core}
The purpose of the resum core is to receive the spikes from all $N/2$ carry-add cores at the same time and output the total number of spikes received by each bit.
This is modeled with $\gamma_2$ neurons.
Since each carry-add operation outputs not more than two spikes per neuron and there are $N/2$ carry-adds, the maximum number of spikes per neuron that the resum cores receives and outputs is $N$.
Here the difference from other cores that use $\gamma_2$ neurons is that not all spikes come at the same tick: the input is coming from carry-add cores which emit spikes during a period of two ticks.
However, each neuron still accumulates the total number of spikes and outputs them correctly since the number of spikes coming to each neuron at the second tick cannot exceed the number of spikes that came at the first tick due to the nature of $\gamma_2$ output from the carry-add cores.

To keep resumming numbers and carrying the overflow bits, the resum core sends the output back into the multiplexer which again sends pairs of numbers into carry-add cores at appropriate clock ticks.

\subsection{The Full Binary Matrix Multiplication Circuit}
\begin{figure}
\includegraphics[width=0.95\columnwidth]{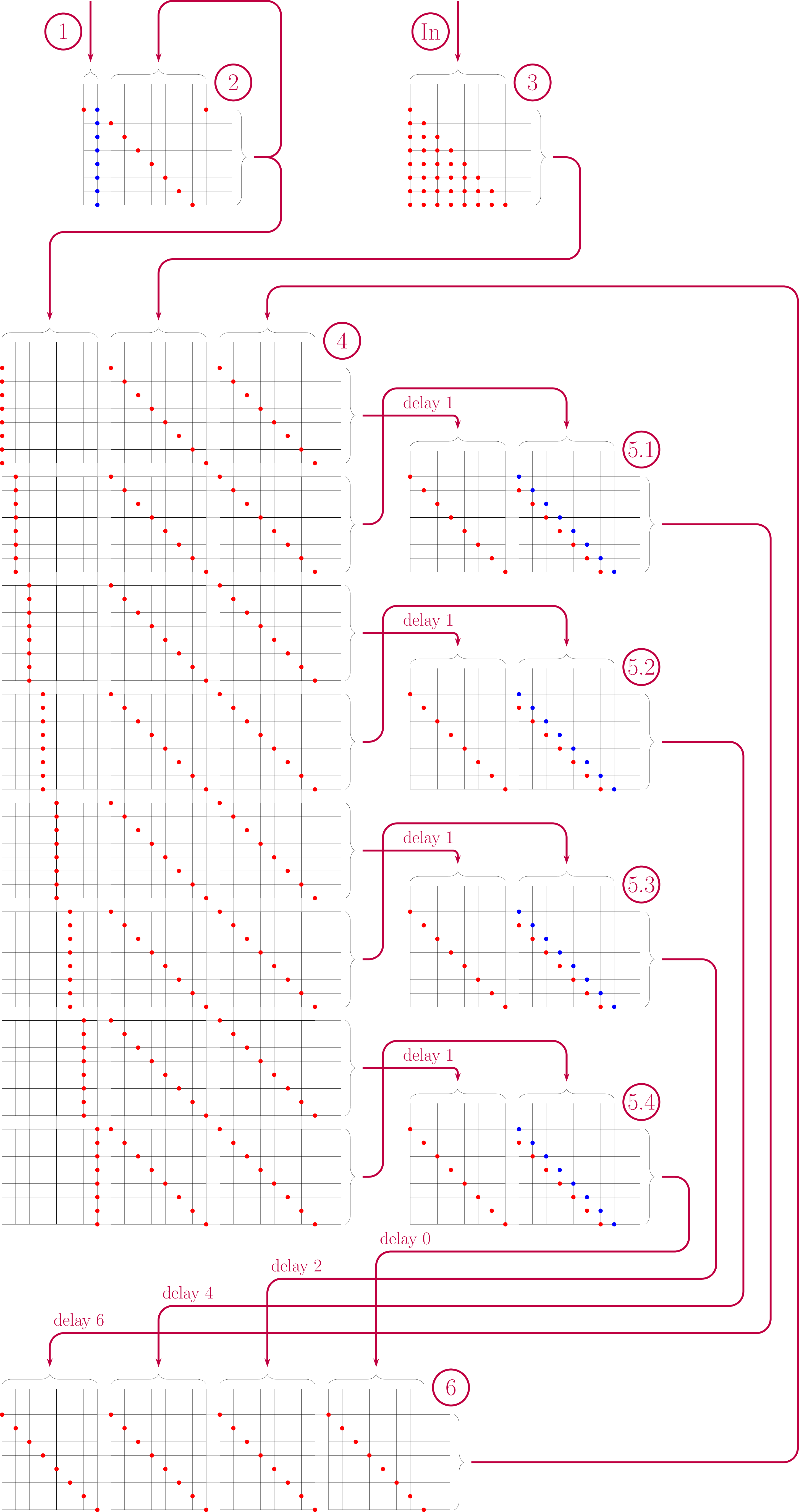}
\caption{The Full Circuit. All cores are shown configured for two's complement multiplication. Note the delays after output from the multiplexer. This coordinates the multiplexer output such that spikes from pairs of registers arrive simultaneously at the carry-add cores. See text for further details.}
\label{fig:circuit}
\end{figure}

A binary multiplication circuit may be constructed by combining the cores described above as shown in Fig.~\ref{fig:circuit}. 
This circuit may be extended to accommodate multiple inputs representing the multiplication of synaptically encoded matrices with spike-encoded input vectors (see Binary Vector-Matrix Multiplication, below). 

The maximum number of clock cycles required for all bits to carry over is $N-1$ with each cycle having $N$ ticks.
The total number of ticks required for the multiplication of two binary numbers is $T = \log X(\log X - 1) + 4$ where $X$ is the resolution of the two's complement numbers.
The total number of neurons used in this circuit is $3\log X + 1.5(\log X)^2$ and the number of synapses is $2\log X + 2(\log X)^2 + 4.5(\log X)^3$.

\subsection{Example Output}

\begin{figure}
\includegraphics[width=0.95\columnwidth]{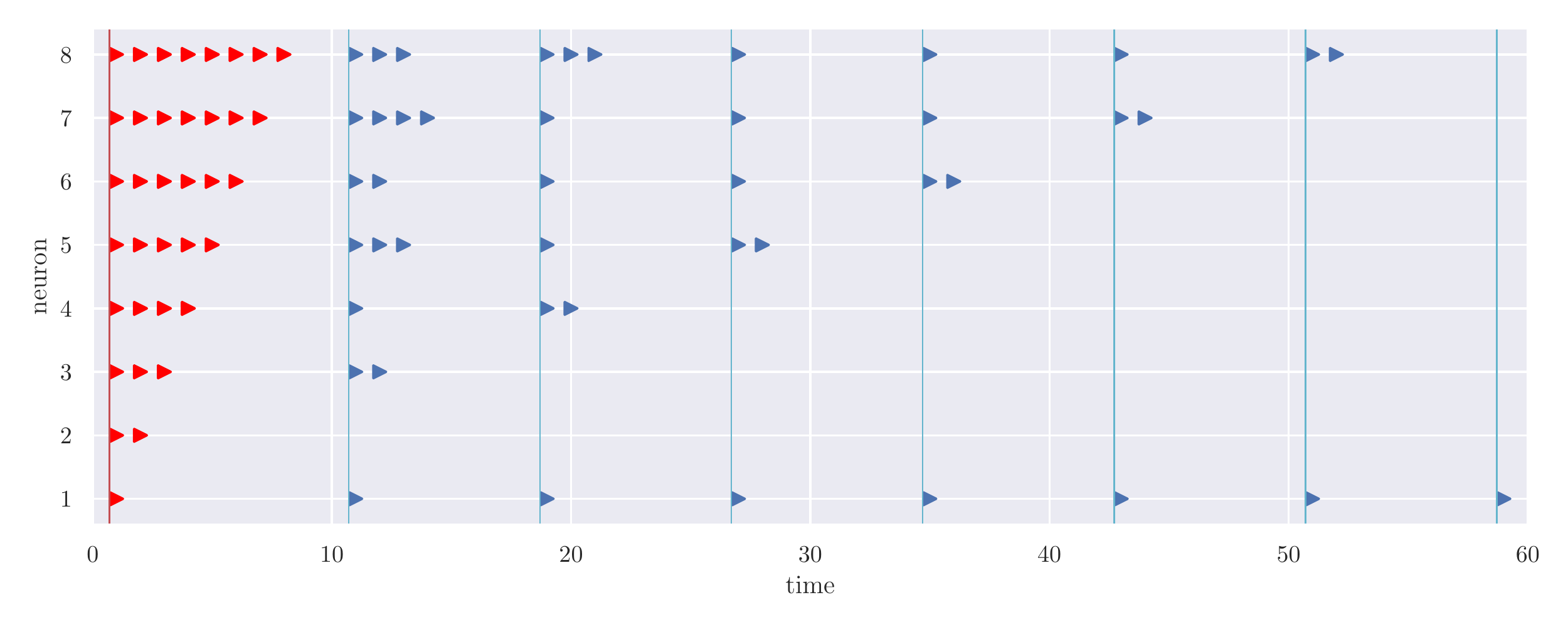}
\caption{An example of the two's complement binary multiplication of $-1\times-1$. Red spikes from the multiply core and blue from the resum core are shown. Successive carry-add operations transform the initial mixed unary-binary encoding into a pure two's complement representation (in this case the number $00000001$.}
\label{fig:bin_mul_example}
\end{figure}

In Fig. \ref{fig:bin_mul_example} we illustrate how the multiply core computes the partial sums that are later summed using carry-add and resum cores for the product $-1\times-1$.
This example is chosen because it has the most nonzero bits in two's complement representation which is $-1 \equiv 11111111$.

The vertical axis represents $8$ bits in multiply and resum cores and the horizontal axis represents time.
The multiply core emits spikes during $8$ ticks which are shown in red.
These sequences of spikes are added in pairs using multiplexer and carry-add cores and sent to the resum core.
The work of the resum core is shown as blue spikes, where the vertical lines indicate the resum cycles.
The resum cycle is shifted compared to the clock since it takes $2$ ticks for spikes to go from the clock to resum core.

All the bits beyond the first one are summed up until eventually the summation in the $8$th bit leads to an overflow and disappearance of spikes.
The result is shown at the beginning of the last resum cycle which represents $2^0 = 1$.

\subsection{Binary Vector-Matrix Multiplication}

\begin{figure}
\includegraphics[width=0.95\columnwidth]{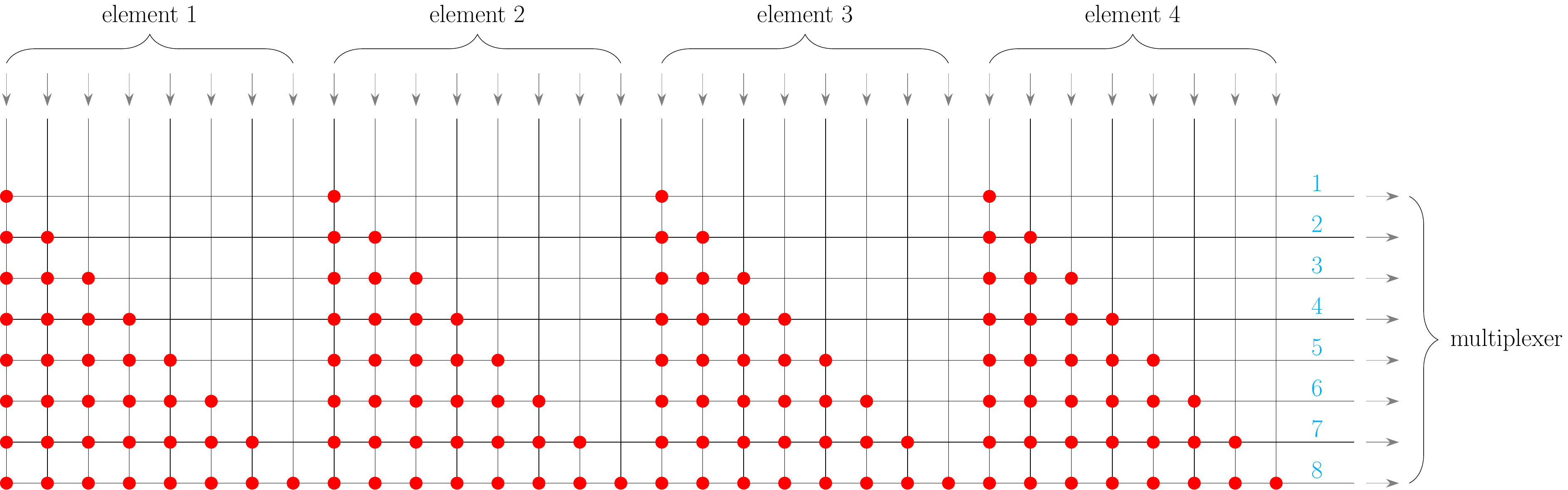}
\caption{Multiply Core for Inner Product. This example circuit shows the two's complement encoding of the vector $(-1, -1, -1, -1)$. To compute an inner product, the elements of the other vector in the inner product are input in the registers indicated as element $i$.}
\label{fig:multiplyvector}
\end{figure}

The two's complement binary multiplication circuit described above can straightforwardly be generalized to multiply $P \times Q$ matrices encoded in synapses with $Q \times 1$ element vectors encoded in spikes. 

To see this, consider the example multiply circuit shown in Fig.~\ref{fig:multiplyvector}. Here, each of the elements of a $4 \times 1$ input vector are first encoded in spikes. They are then input to a multiply core that has been extended to compute the inner product of one row of the matrix with the input vector.

This results in a similar spiking output to that of the above circuit. The main difference is that many more spikes can potentially be output due to the multiple terms in the inner product. Nonetheless, with the same cores described above, the full computation can be performed for all rows of the matrix of interest, resulting in a binary representation of the vector-matrix computation output.

A na\"ive implementation of the above concept results in a circuit with a total number of ticks of $T = Q \log X (Q \log X - 1) + 4$, total neurons of $3 Q \log X + 1.5 (Q \log X)^2$, and total synapses of $2 Q \log X + 2 (Q \log X)^2 + 4.5(Q \log X)^3$. However, by incorporating a second set of clock, multiplexer, carry-add, and resum cores, the time may be reduced at the expense of using more neurons and synapses.

To do this, the first clock performs a first carry-add cycle, but then triggers the second clock, which has a cycle length half that of the first clock. Spikes output from the first resum core are propagated to the second multiplexer, which only takes inputs for half the time of the first clock (as the initial carry-adds decrease the number of spikes considerably) and outputs half the number of carry-add pairs as the first multiplexer. This significantly decreases the time needed for the full multiplication procedure to complete.

More complex carry-add cycle management could further reduce the time needed for a matrix-vector multiplication procedure, but we have not investigated further refinements in detail.

With this matrix multiplication circuit in hand, one can then integrate discretized partial differential equations, deep nets, or other linear algebraic operations in the usual way. For instance, the diffusion equation may be discretized to a set of ordinary differential equations. The time evolution is then computed iteratively via matrix multiplication.

\subsection{Random Walk Circuit}
In this section, we present a circuit which computes a 1-dimensional random walk that starts at a location $x_0$ and chooses to step $\delta_\pm = \pm1$ with probability $p_\pm = 0.5$.
We again use two's complement encoding for N-bit numbers.
For independent random walkers, the full multiplication circuit is unnecessary but carry-add operations are still necessary. This allows for a significant reduction in circuit size.
We introduce new types of neurons that we use in this circuit:
\begin{itemize}
    \item $\gamma_3$ -- a CONSTANT neuron, which emits spikes every $M$th tick;
    \item $\gamma_4$ -- a RANDOM neuron, which emits spikes every tick with probability $p = 0.5$.
\end{itemize}
The behavior of the $\gamma_3$ neuron is straightforward and is modeled with the help of bias current and refractory delay.
The period $M$ is equal to the length of a single summation cycle which is $N+1$.
In order to model randomness we use Loihi's capability to add random noise to the voltage threshold.
This is implemented by setting noise to take only two values (0 and 2) and bias current to be equal to the voltage threshold.
Thus the total voltage exceeds the threshold with probability $0.5$ as needed.

The full circuit consists of two single neurons, $\gamma_3$ (constant) and $\gamma_4$ (random), and three cores: a perturbation core of $\gamma_1$ neurons, a location core of $\gamma_0$ neurons, and a carry core of $\gamma_1$ neurons.
The circuit needs $N+1$ ticks after supplying the perturbation value $\pm1$ to the location core to perform the summation operation.
If an overflow happens in the first digit, it can need up to $N$ more operations to carry the overflow to more significant digits.

\begin{figure}
\includegraphics[width=0.45\columnwidth]{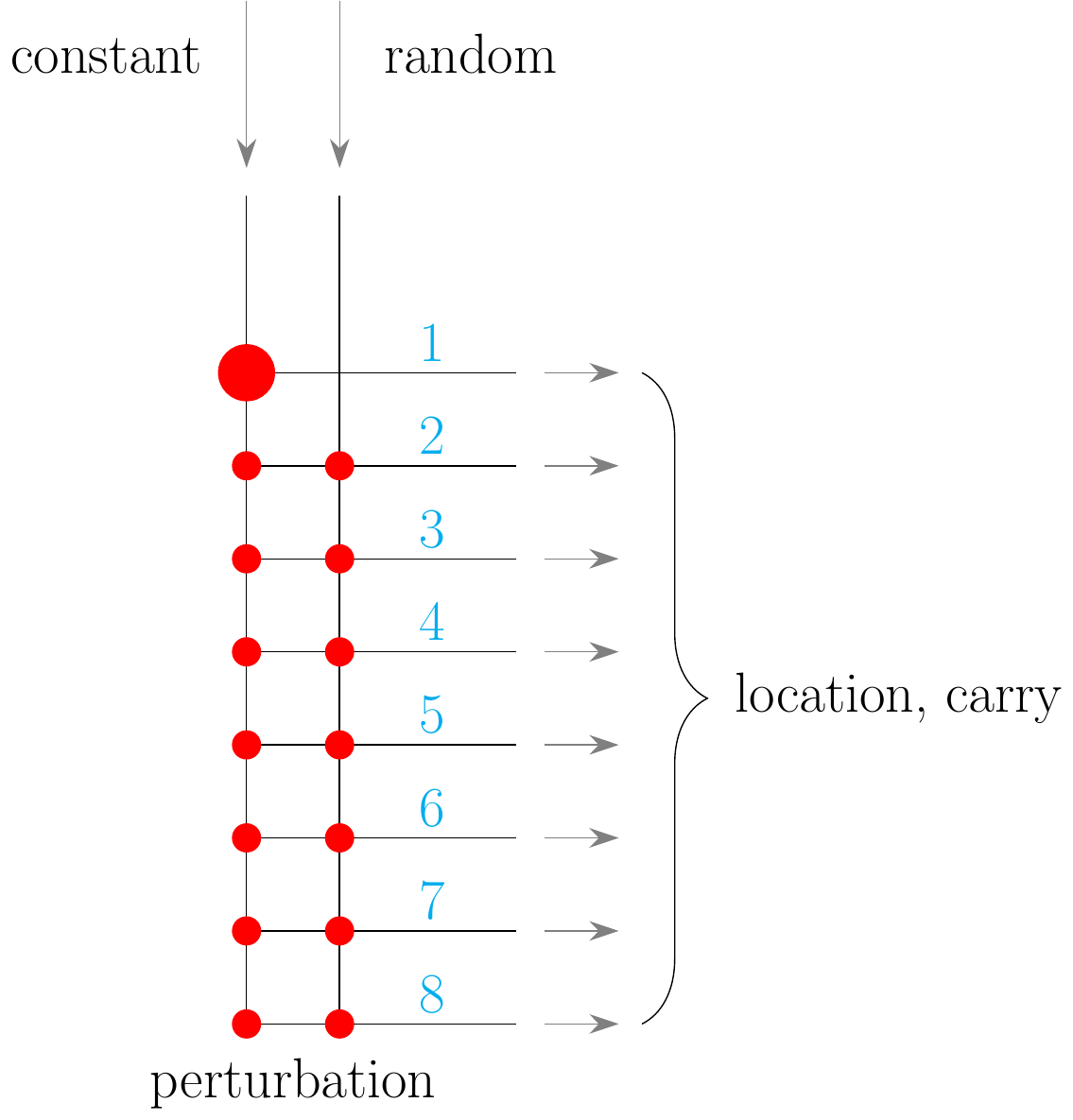}
\caption{The Perturbation Core. Here a spike is input at every iteration (constant) and a spike is randomly input at every iteration. In concert with the given synapses, an output of either $+1$ or $-1$ is output. The output drives the downstream location and carry circuits in the random walker.}
\label{fig:perturbation}
\end{figure}

\begin{figure}
\includegraphics[width=0.95\columnwidth]{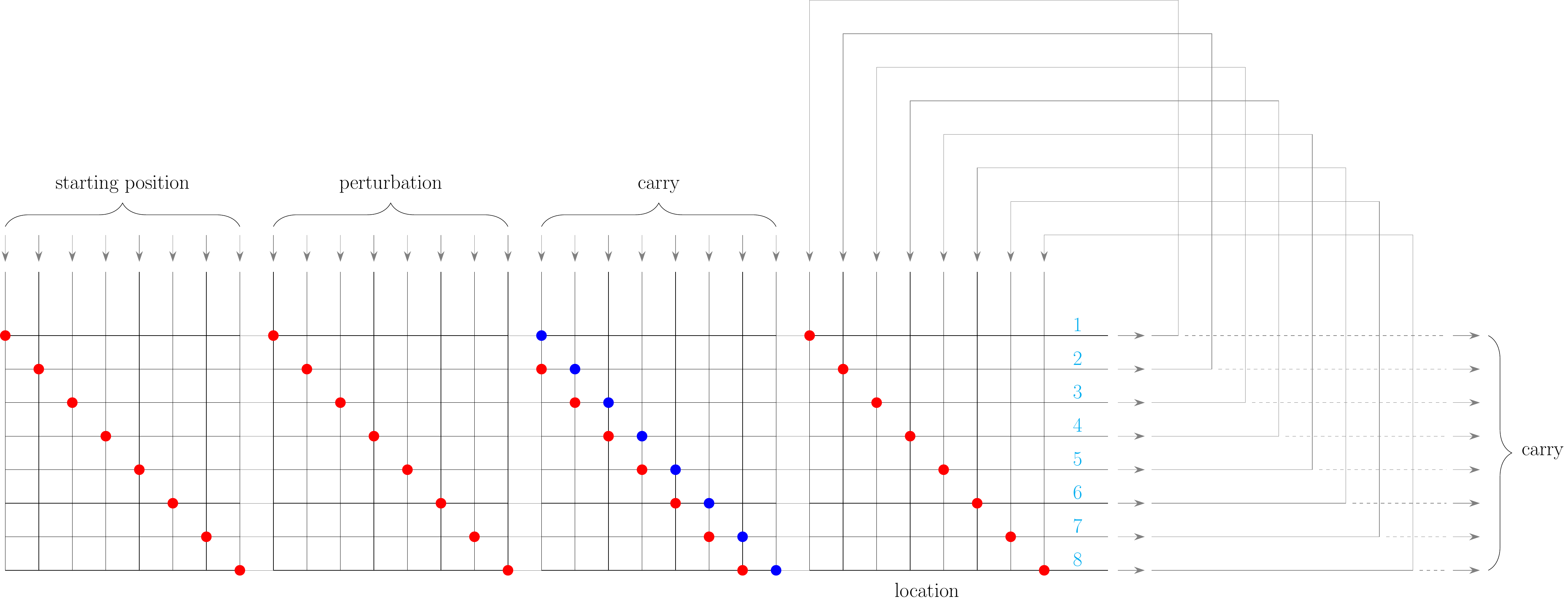}
\caption{The Location Core. A starting location (used only once for initialization), perturbation from the perturbation core, and carry from the carry core are input. The carry is output and also fed back to the location register.}
\label{fig:location}
\end{figure}

\begin{figure}
\includegraphics[width=0.95\columnwidth]{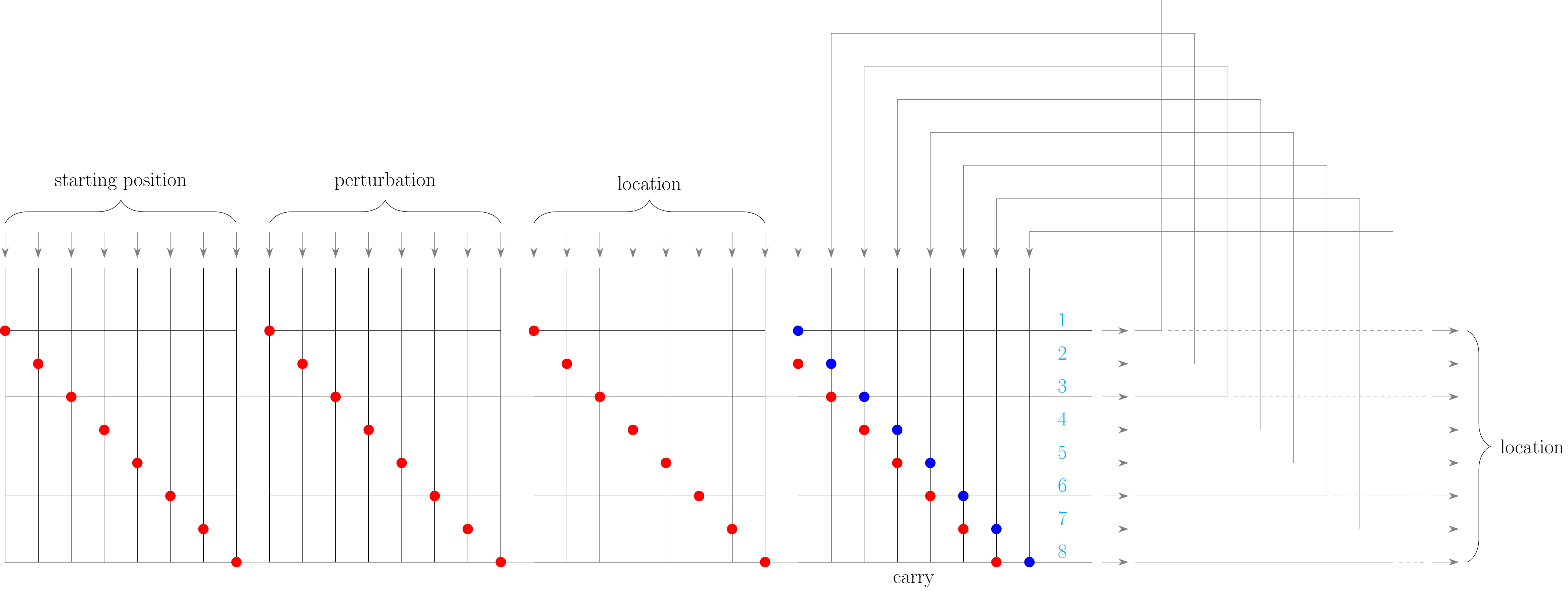}
\caption{The Carry Core. This core has the same structure as the location core, but with location and carry registers reversed.}
\label{fig:carry}
\end{figure}

The diagrams in Figs.~\ref{fig:perturbation}, \ref{fig:location}, and \ref{fig:carry} show  the circuit elements.
The synaptic weights in the perturbation core combine the spikes coming from neurons $\gamma_3$ and $\gamma_4$ in order to output the binary representation of either $+1$ or $-1$ depending on the output of $\gamma_4$ (Figure \ref{fig:perturbation}).
The large red circle in the first neuron represents a double weight which is needed to pass the larger threshold of $\gamma_1$.

The location and carry cores receive the same input (starting position, perturbation, carry, and location) but process it with different neurons: $\gamma_0$ for the location core and $\gamma_1$ for the carry core (Figures \ref{fig:location} and \ref{fig:carry}).
The circuit works in such a way that at each tick these two cores receive spikes from not more than 2 inputs (either starting position and perturbation, location and perturbation, or location and carry).
The sum of location and carry represents the current location and it can take up to $N$ ticks to carry the overflow digits after receiving the perturbation spikes.
After $N$ ticks the carry stops producing any spikes and the location core holds the complete current location result.

The number of ticks required for this circuit to make $M$ steps in terms of resolution is $T = 2 + M(\log X + 1)$.
The number of neurons used for a single agent is $2 + 3\log X$ and the number of synapses is $2\log X + 8(\log X)^2$.

\begin{figure}
\includegraphics[width=0.95\columnwidth]{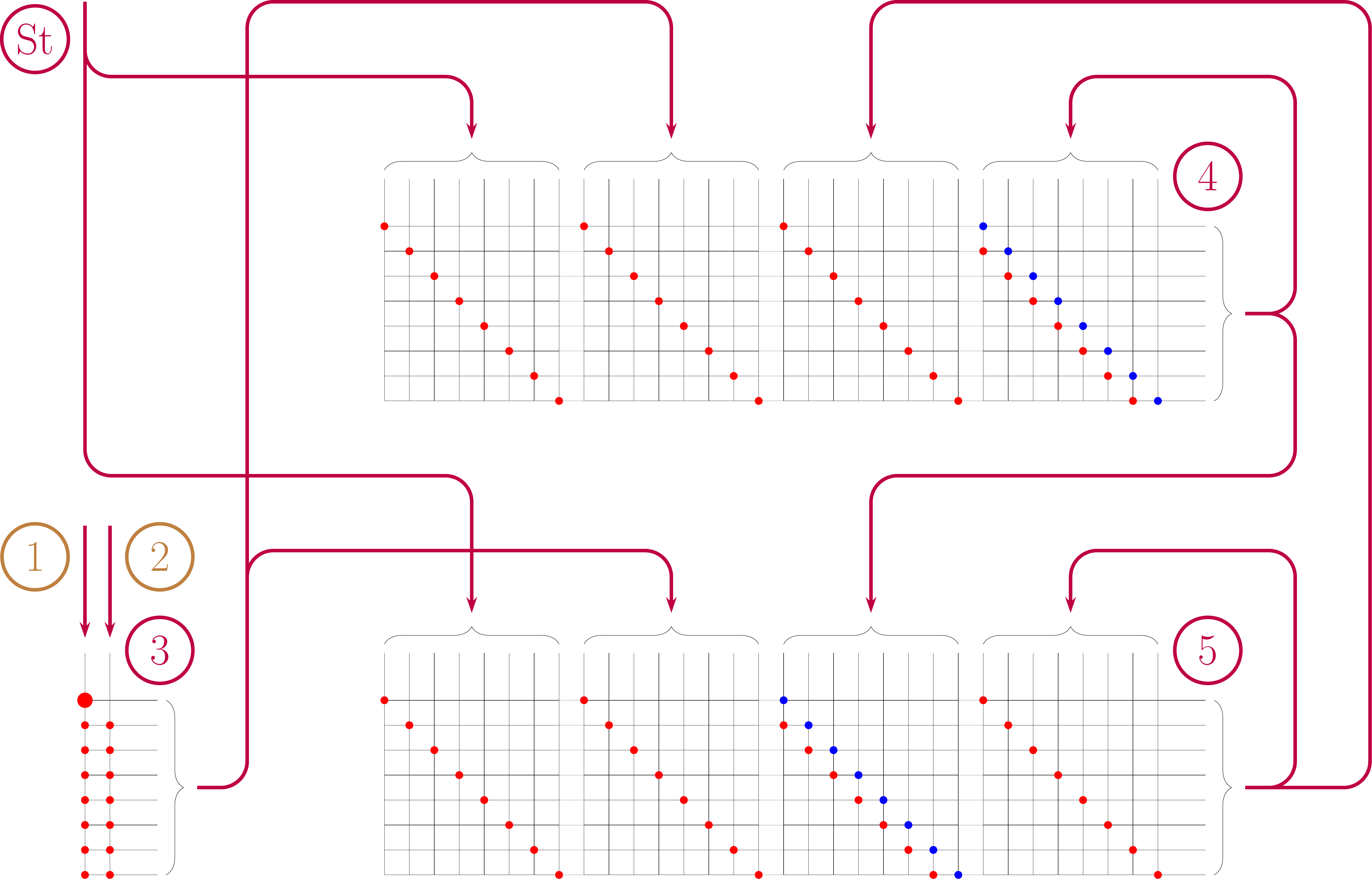}
\caption{The Random Walk Circuit. This simple circuit combines the perturbation, location, and carry cores to perform a random walk. See text for details.}
\label{fig:random_walk}
\end{figure}

The full circuit is depicted on Figure \ref{fig:random_walk}.

\section{Results}

\begin{figure}
\includegraphics[width=0.95\columnwidth]{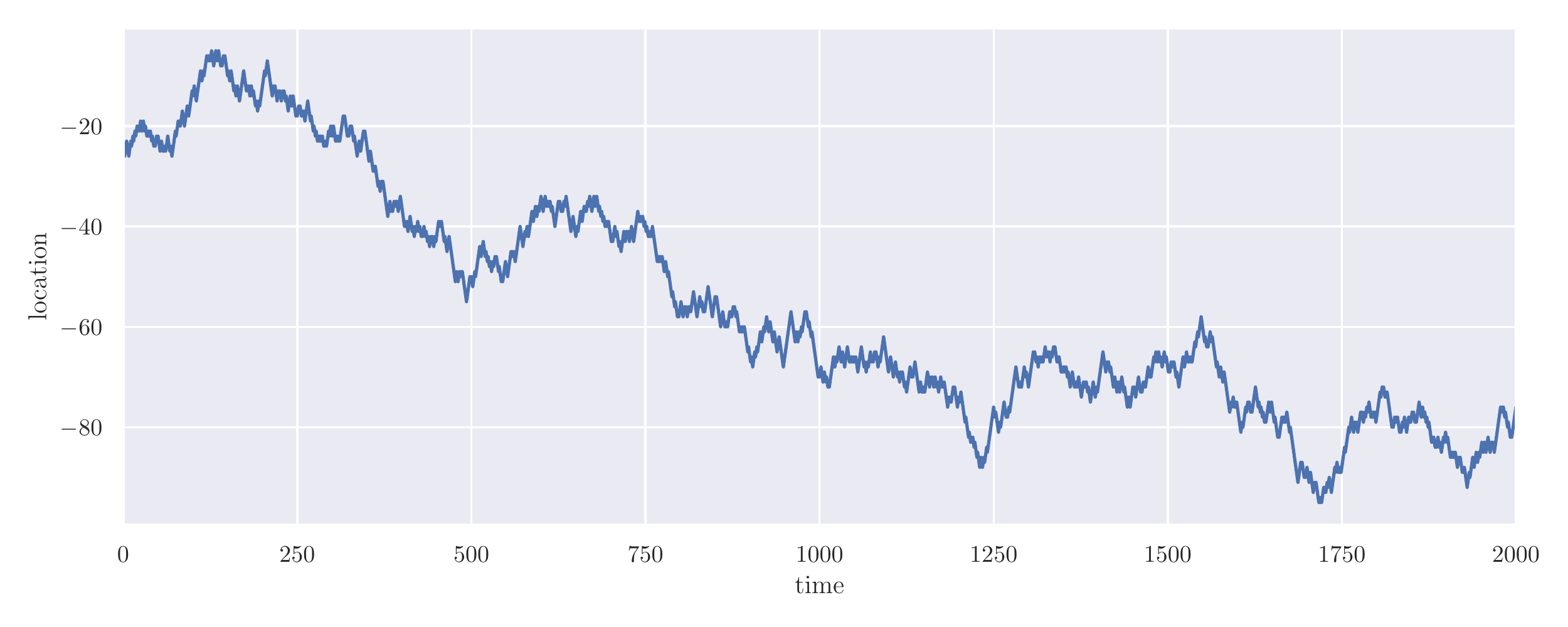}
\caption{The random walk of a single agent in $8$-bit resolution as a function of time.}
\label{fig:single_walk}
\end{figure}

\begin{figure}
\includegraphics[width=0.95\columnwidth]{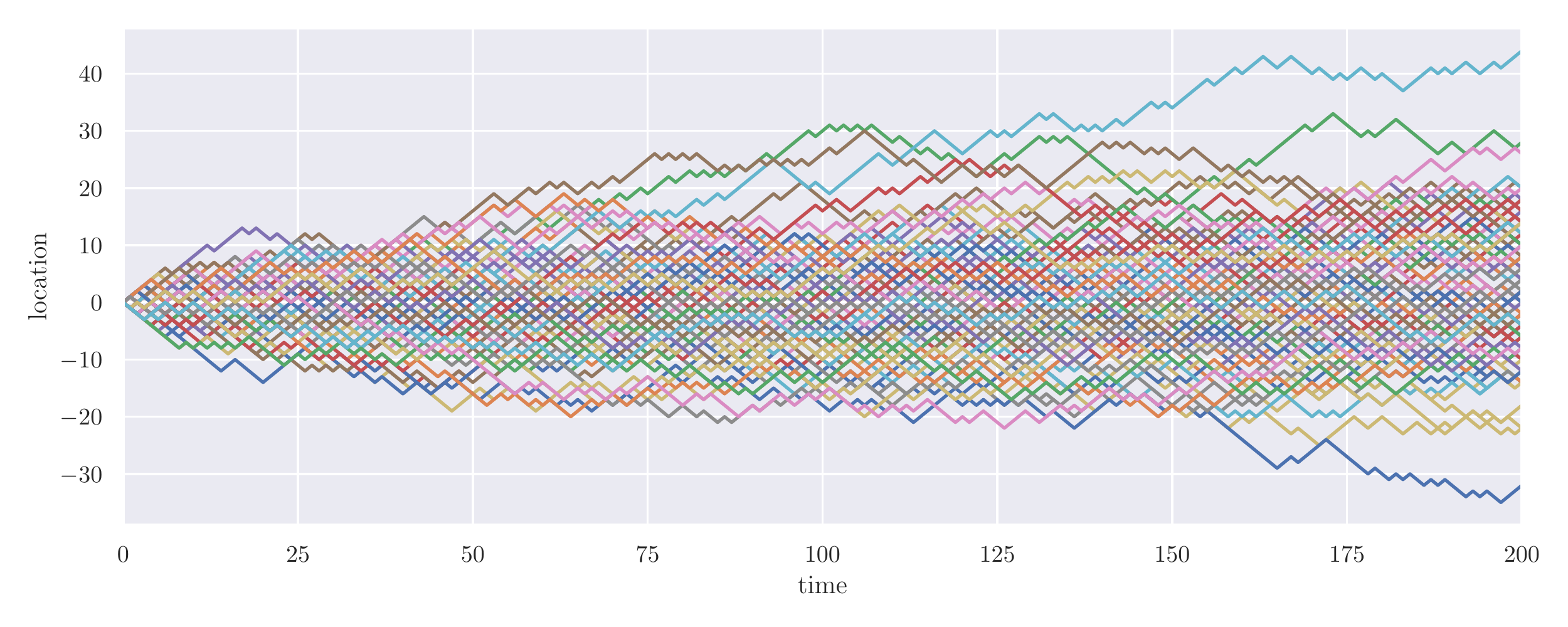}
\caption{The random walks of several agents in $8$-bit resolution as a function of time.}
\label{fig:multi_walk}
\end{figure}

\begin{figure}
\includegraphics[width=0.95\columnwidth]{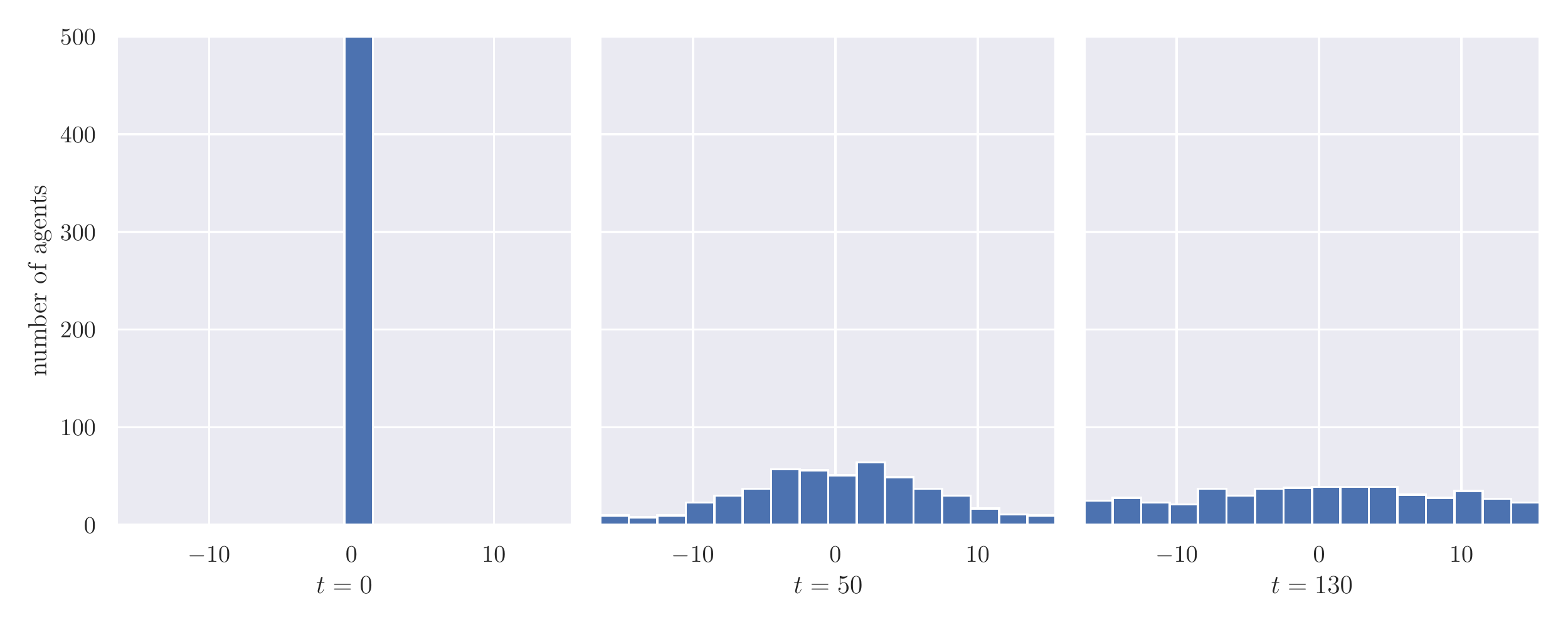}
\caption{The distributions of agents' locations as the distribution evolves from a delta function, to a normal distribution, to an asymptotically flat distribution.}
\label{fig:random_walk_dist}
\end{figure}

Fig. \ref{fig:single_walk} shows the random walk of a single agent in $8$-bit space.
Fig. \ref{fig:multi_walk} shows diffusion by several agents in $8$-bit space. 
Fig. \ref{fig:random_walk_dist} shows the distribution of the locations of $500$ agents at times $t = 0, 50, 130$ in $5$-bit space.
Here the space was taken small to illustrate the effect of the periodic boundary conditions which eventually lead from a Gaussian to uniform distribution.

\begin{figure}
\includegraphics[width=0.95\columnwidth]{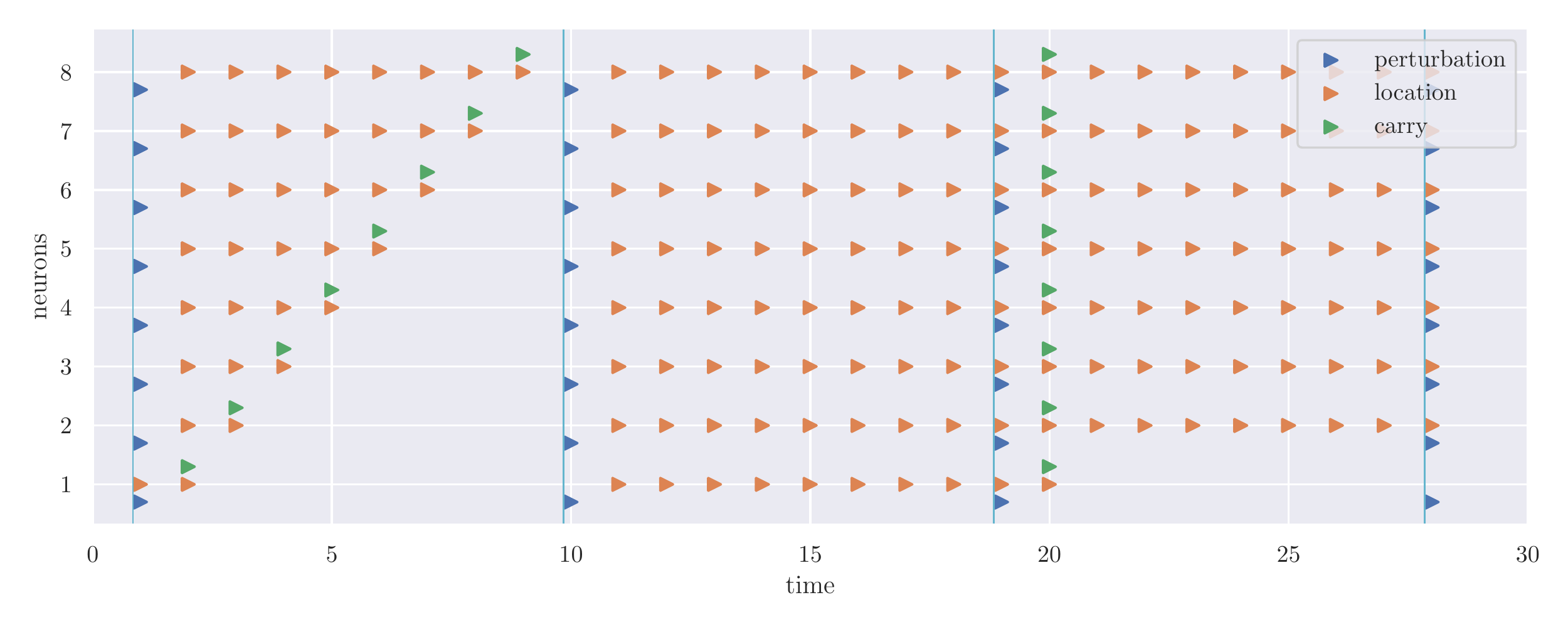}
\caption{Illustration of spikes in a random walk scheme that starts at $x_0 = 1$ and makes three steps, each time receiving a perturbation of $-1$. Spikes from all three cores are shown superposed.}
\label{fig:random_walk_example-2}
\end{figure}

In Fig.~\ref{fig:random_walk_example-2}, we show spike propagation as a function of time in perturbation, location, and carry cores superimposed. Here, three repeated perturbations of $-1$ increment the location from an initial value of $0$, successively to $-1$, then $-2$ (two's complement). The action of the location and carry cores is evident as the spikes are successively reduced to final two's complement numbers at each iteration.

\begin{figure}
\includegraphics[width=0.95\columnwidth]{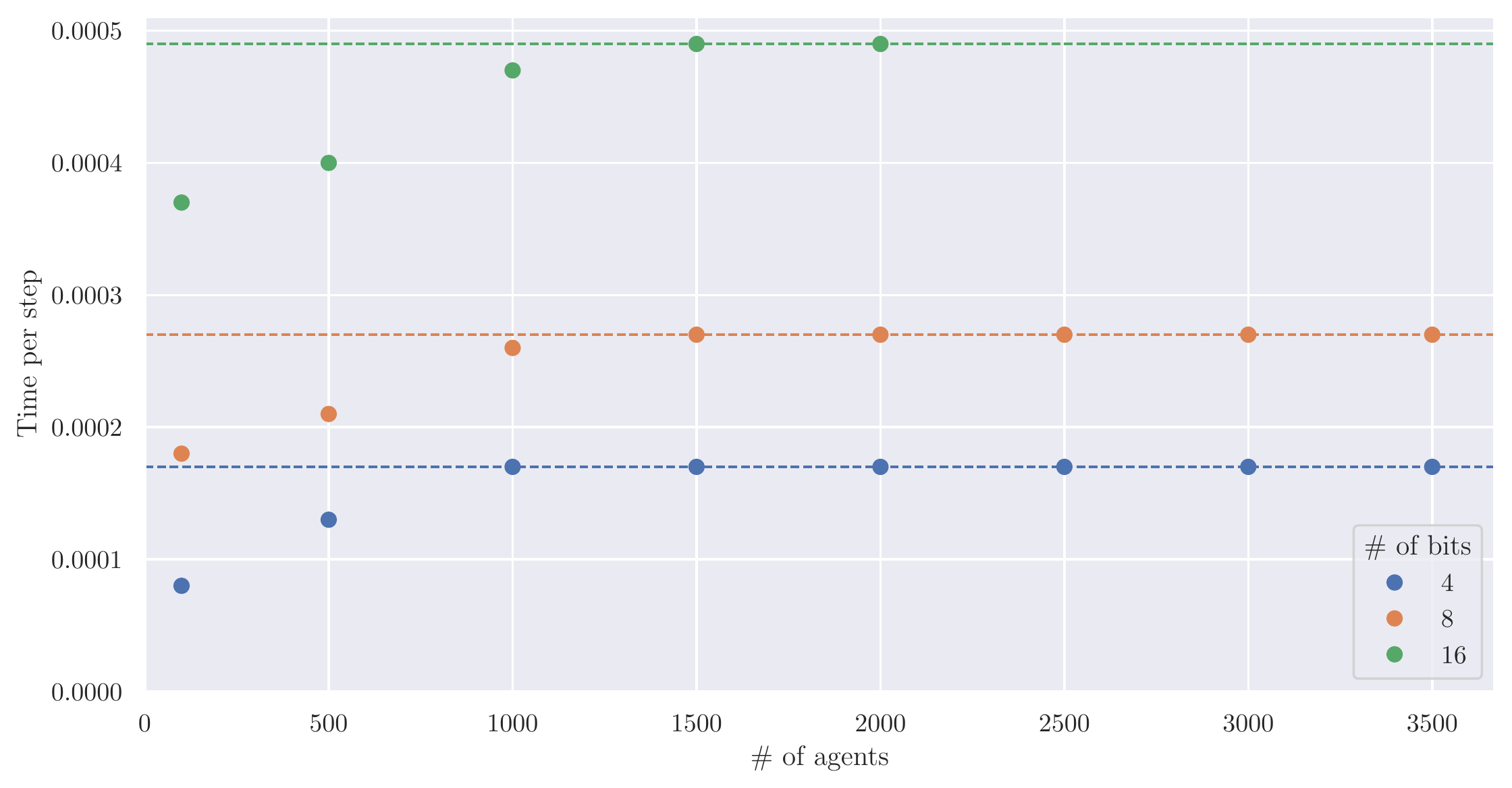}
\caption{The performance of the random walk (time per random walk step) algorithm implemented on Loihi as a function of the number of agents distributed on the chip. Note that, at a certain point, the number of $16$-bit walkers exceeded the size of our Kapoho Bay system, so we were only able to assess up to $2000$ $16$-bit walkers.}
\label{fig:scaling}
\end{figure}

The performance of the random walk circuit on Loihi is shown in Figure \ref{fig:scaling} for different resolutions and numbers of agents.
All runs contain $1000$ steps and we expect the performance to be independent of the number of agents since Loihi propagates all agents in parallel.
We see that the time per tick is almost independent of the resolution of the number which results in time per random walk step being proportional to $\log X$ where $X$ is the resolution.

However, the plot shows an initial linear growth of time with the number of agents. This indicates the counterintuitive result that only asymptotically is good parallelization achieved.
Without access to the full scheduling and distribution of the circuit on chip it is difficult to pinpoint the reason for the initial growth in iteration time versus number of agents.


\section{Discussion}
\noindent
We have constructed a neuromorphic circuit that performs a matrix-vector product in two's complement representation. The timing of the computation is controlled by a clock, and a set of other cores is used to carry out the reduction of the representation to a final, two's complement encoded binary integer.

We have shown how this circuit can be extended for application to matrix multiplication and hence used for various applications, such as the integration of linear equations.

Additionally, we introduced a simpler circuit, still using two's complement numbers to perform a random walk and shown that this circuit parallelizes well and asymptotes to a constant time per step for large numbers of agents.

Our work here demonstrates that, at the cost of a larger neuronal footprint, binary coded multiplication may be performed on circuits with sizes and run times that are poly-logarithmic in the resolution of the numbers being multiplied. This results in a significant savings in multiplication time and spikes generated relative to unary coding, where number resolution is proportional to spike number. This can result in significant reductions in power consumed due to the relationship between spike number and power consumption.


\begin{acknowledgments}
{\bf Acknowledgements}. We thank Alpha Renner for helpful conversations. We thank Intel's Loihi development team and the Intel Neuromorphic Research Community for their support. This work was supported by the LANL ASC Beyond Moore's Law project (O.I.) and by the US Department of Energy National Nuclear Security Administration’s Office of Defense Nuclear Nonproliferation Research \& Development (DNN R\&D) at Los Alamos National Laboratory under contract 89233218CNA000001 (A.T.S.). LANL approval designation: LA-UR-21-22286
\end{acknowledgments}

\bibliographystyle{unsrtnat}
\bibliography{Biblio}

\end{document}